\documentstyle[12pt,aps]{revtex}

\begin{document}
\begin{center}
\begin{Large}
Hypersensitive transport in a phase model with multiplicative
stimulus
\end{Large}
\bigskip

Saul.L.Ginzburg and Mark.A.Pustovoit

\bigskip

B.P.Konstantinov Petersburg Nuclear Physics Institute, Gatchina
Leningrad district 188350 Russia
\bigskip

Abstract
\end{center}

 In a simple system with periodic symmetric
potential, the phase model under effect of strong multiplicative
noise or periodic square wave, we found a giant response, in the
form of directed flux, to an ultrasmall dc signal. The resulting
flux demonstrates a bell-shaped dependence on multiplicative noise
correlation time and occurs even in the case of large (compared to
the signal) additive noise.
\bigskip


\bigskip

The phenomenon of noise-induced transport, when a directed
transport of matter emerges in a stochastic system with periodic
potential in the absence of directed external force, attracts now
a vivid and stable interest \cite{reviews}. An active theoretical
and numerical investigations of such model systems with asymmetric
potential (``thermal ratchets'') was stimulated by their potential
ability to explain functioning of motor proteins, which are
responsible for cell motion. However, the applicability of the
concept is more wide. Several experimental groups found
rectification of Brownian movement in chemical systems (directed
diffusion (\cite{exper1},\cite{exper2}) or rotation
(\cite{exper3},\cite{exper4}) of molecules), therefore pointing a
possible application of ``ratchet effect'' in nanoscale mechanical
devices and separation technique.

Noise-induced transport was also found in systems with symmetric
periodic potential (the term ``ratchet'' is confusing in this
case) (\cite{millonas1}--\cite{berdic2}). The mostly wide known of
models for such systems is the phase model $d\varphi/dt = a -
b\sin \varphi$, which has been applied for the study of various
physical systems, $e.g.$, superionic conductors, Josephson
junction arrays, chemical reactions, ring-laser gyroscopes -- up
to neural networks. The good review of physical applications of
phase model can be found in \cite{risken}. The model demonstrates,
under stochastic stimulus, numerous interesting phenomena
including stochastic resonance and noise-induced transport
(\cite{phasemodel}-\cite{berdic2}).

The attention of investigators of noise-induced transport was,
until now, focused mainly on directed flux of particles induced by
noise \textit{with zero mean, in the absence of constant external
driving}. In the present work we consider \textbf{\textit{the
response to an ultrasmall external signal}} in phase model with
strong symmetric multiplicative noise. We show that a macroscopic
flux of matter appears in such a system under effect of ultrasmall
dc driving. This effect resembles the phenomenon of noise-induced
hypersensitivity to small signals recently found by us
\cite{ours}, and therefore we call the new phenomenon as
\textbf{\textit{hypersensitive transport}}\textit{ (HST).}

We study the phenomenon of HST for telegraph (dichotomous) and
Gaussian colored noise, and observe HST in a wide range of noise
correlation times, with bell-shaped dependence of transport
velocity (voltage) on the latter . Also we consider the case of
entirely deterministic system when a multiplicative periodic
stimulus replaces a multiplicative noise; HST in this case occurs
as well.

The system under study is described by the following dimensionless
equation:

\begin{equation}
\label{eq1} {\frac{{d\varphi} }{{dt}}} = {\left[ {z\left( {t}
\right) - z_{c}} \right]}\sin \varphi + F + \sigma \psi \left( {t}
\right){\rm .}
\end{equation}

Here $z_{c},  \sigma $ are the positive constants, and $\psi (t)$
is $\delta $-correlated white noise source. The additive dc term
$F$ is called further the signal. The control stimulus $z(t)$ can
be stochastic or deterministic function. We consider here several
types of it, namely, exponentially correlated dichotomous noise
with amplitude $\Delta $ and flipping rate $\gamma /2$, so that
${\left\langle {z\left( {t} \right)z\left( {{t}'} \right)}
\right\rangle}  = \Delta ^{2}\exp {\left[ { - \gamma \left( {t -
{t}'} \right)} \right]}$, colored Gaussian noise with the same
autocorrelator, and deterministic square wave with amplitude
$\Delta $ and period $T$. We restrict ourselves by the case of
small signal, $i.e.$, $z_{c} ,\,\Delta ,\,\gamma ,\,T^{ - 1} \gg
\sigma ,\,F\sim 10^{ - n},\quad n \gg 1$.

The problem (\ref{eq1}) was considered by several authors
(\cite{millonas2},\cite{phasemodel}-\cite{berdic2}). However,
nobody of them paid a special attention to the region of
ultrasmall $\sigma $ and $F$, where the phenomenon of HST occurs.

First, we note that we use sufficiently large stimulus ($\Delta >
z_{c} $), so when it is switched, the stable and unstable fixed
points in Eq. (\ref{eq1}) are switched also. In fact, Eq.
(\ref{eq1}) at different time moments becomes:

\begin{eqnarray}
\label{eq2}
&{\frac{{d\varphi} }{{dt}}} = f_{\pm}  \left(
{\varphi}\right) + F + \sigma \psi (t), \nonumber \\
&f_{\pm}  \left( {\varphi} \right) = \left( {\pm \Delta - z_{c}}
\right)\sin \varphi = - {\frac{{\partial U_{\pm}  \left( {\varphi}
\right)}}{{\partial \varphi} }}{\rm .}
\end{eqnarray}

The stable fixed points (FPs) $\varphi _{n +} ^{s} $, $\varphi _{n
-} ^{s} $ and the unstable ones $\varphi _{n +} ^{u} $, $\varphi
_{n -} ^{u} $ of Eq.(\ref{eq2}) are:

\begin{eqnarray}
 &\varphi _{n + }^s  = \pi \left( {2n + 1} \right) + \frac{F}{{\Delta  - z_c }},
 \quad \varphi _{n - }^s  = 2\pi n + \frac{F}{{\Delta  + z_c }},
 \nonumber \\
 &\varphi _{n + }^u  = 2\pi n - \frac{F}{{\Delta  - z_c }},
 \quad \varphi _{n - }^u  = \pi \left( {2n + 1} \right) - \frac{F}{{\Delta  + z_c }}.
 \label{eq3}
\end{eqnarray}

These FPs together with the potential $U_{\pm}  \left( {\varphi}
\right) + F$ are depicted in Fig.1. One can see that for small $F$
the unstable FPs of one branch are located left, and very close
to, the stable FPs of other branch as well as to the points $\pi
n$.

We point out now an interesting feature of the system that results
in appearance of HST. We restrict ourselves in further sketch by
the case of dichotomous noise.

Let us suppose that $z(t)=\Delta $ and the system is in the stable
FP $\varphi _{0 +} ^{s} $. When $z(t)$ changes its sign, the
system turns to the branch $f_{ -} +F$, therefore moving right
(Fig.1). For sufficiently rarely switching stimulus ($\Delta \gg
\gamma  ,T^{-1}$) the system passes through subsequent unstable
point $\varphi _{1 +} ^{u} $ before stimulus is switched again. As
a result, macroscopic transport appears. Such a picture was
described in \cite{berdic2}, however they did not considered the
possibility of macroscopic response to an ultrasmall driving. For
frequently switching stimulus, when the condition $\Delta  >
\gamma $ is not valid, the system may not have enough time to
reach $\varphi _{1 +} ^{u} $, and after subsequent switching of
$z(t)$ it will turn to the branch $f_{ +} +F$ and will move left.
The system obviously can not move left of $\varphi _{0 +} ^{s} $,
thus the latter point is a reflecting boundary for all $\varphi $
in the interval $\left( {\varphi _{0 + }^{s} ,\varphi _{1 +} ^{u}}
\right)$. For square wave stimulus the system enters a closed
loop, with any transport disappearing. But for stochastic $z(t)$,
\textit{even when it flips rapidly}, there is nonzero probability
of passage of the system through the point $\varphi _{1 +} ^{u} $
along the branch $f_{ -} +F$. After that the system can not go
back left, but easily will go right through $\varphi _{1 -} ^{s} $
along the branch $f_{ +} +F$ after next switching of stimulus.
Thus we see that the intervals [$\varphi _{n -} ^{u} ,\varphi _{n
+} ^{s} $] and [$\varphi _{n +} ^{u} ,\varphi _{n -} ^{s} $] act
as \textit{semitransparent mirrors}, passing the system through
only to the right. The width of these intervals is $2F\Delta
\left( {\Delta ^{2} - z_{c}^{2}}  \right)^{ - 1}\sim 10^{ - n} \ll
1$. The described picture is valid only in the absence of additive
noise. Otherwise, the system is able to pass through mirrors in
both directions, however moving right predominantly, and HST still
occurs.

Now let us obtain an expression for transport velocity for
adiabatic stimulus. In this case, the system should come to
equilibrium during the characteristic time of the process ($T$ or
$1/\gamma $). This means that the system variable $\varphi $
should decrease to the value of the order of $F$ or $\sigma $,
having the relaxation speed of the order of $\Delta - z_{c}$.
Mathematically, this statement can be expressed as: $ \exp \left(
-\left( \Delta - z_{c} \right) T \right) \ll F,\quad
 \exp \left( -\left( \Delta - z_{c} \right)\gamma^{-1} \right) \ll F$.
Thus we get the following criteria:

\begin{equation}
\label{eq4}
 T^{ - 1},\gamma \ll {\frac{\Delta - z_{c}} {\ln 1/F}}
\sim {\frac{\Delta - z_{c} }{n}}.
\end{equation}

From Eq.(\ref{eq4}) we see that the macroscopic transport can
occur in adiabatic regime at not-so-small stimulus frequency even
for ultrasmall $F$.

Let us suppose the system to be at ``positive'' branch
(\textit{z(t)=$\Delta $}) near stable FP, say $\varphi _{0 +} ^{s}
$, in the local equilibrium state. The probability density of
\textit{$\varphi $} in this case is $P_{ +}  \left( {\varphi}
\right) = C\exp \left( { - {\textstyle{{2U_{ +}  \left( {\varphi}
\right)} \over {\sigma ^{2}}}}} \right)$($C$ is the normalization
constant). When $z(t)$ is switched and the system turns to
``negative'' branch, we see from Fig.1 that, if the system was
located right of the point $\varphi _{0 -} ^{u} $, it begins to
move right and after some time reaches $\varphi _{1 -} ^{s} $,
increasing \textit{$\varphi $} by \textit{$\pi $}. The probability
of this event is $w_{1}^{ +}  = {\int\limits_{\pi - {{F} \over
{\Delta+z_{c}}}}^{\infty}  {P_{ +}  \left( {\varphi} \right)}}
d\varphi $. If the system in the moment of switching appeared at
the left of $\varphi _{0 - }^{u} $it will move left up to the
point $\varphi _{0 -} ^{s} $(with \textit{$\varphi $} decrease by
\textit{$\pi $}) with the probability $w_{1}^{ -}  =
{\int\limits_{ - \infty }^{\pi - {{F} \over {\Delta+z_{c}}}} {P_{
+} \left( {\varphi} \right)}} d\varphi $. If the system starts
from ``negative'' branch, we obtain the transport probabilities
$w^{\pm} _{2}$ in the same way. After calculation of integrals we
have:

\begin{eqnarray}
 &w_{1,2}^{\pm}  = {\textstyle{{1} \over {2}}}\left(
{1\pm \Phi \left( {a_{1,2}}  \right)} \right),
\nonumber \\
&a_{1,2} \equiv {\frac{{2F\Delta} }{{\sigma \left( {\Delta \pm
z_{c}} \right)\sqrt {\Delta \mp z_{c}} } }}\;, &\Phi \left( {z}
\right) = {\frac{{2}}{{\sqrt {\pi} } }}{\int\limits_{0}^{z} {e^{ -
z^{2}}dz\;.}} \label{eq5}
\end{eqnarray}

Now we see that for $\Delta \pm  z_{c}\sim 1, \quad F, \sigma \sim
 10^{-n}$ the transport probabilities
$w^{\pm} _{1,2}$ depend solely on the ratio $F/\sigma $. In
zero-temperature case, when the additive noise is absent, the
probabilities to move right become equal to unity.

Let us note now that in adiabatic case the system spends most of
the time near stable FPs, hopping between them with an average
rate $\gamma /2$ (or $2/T$ for square-wave stimulus), and we can
discretize the dynamic variable: $\varphi \left( {t} \right) = \pi
n\left( {t} \right)$. Solving the corresponding master equation,
we obtain the following expression for transport velocity $V =
{\left\langle d\varphi/dt\right\rangle} $ for stochastic stimulus:

\begin{eqnarray}
 &V = {{{\pi \gamma}  \over {4}}}\left( {w_{1}^{ +}  + w_{2}^{ +}
- w_{1}^{ -}  - w_{2}^{ -} }  \right) = {\textstyle{{\pi \gamma}
\over {4}}}\left( {\Phi \left( {a_{1}}  \right) + \Phi \left(
{a_{2}}  \right)} \right),
\nonumber \\
 &V = {\textstyle{{\pi \gamma}  \over {2}}},\quad \sigma = 0\;.
\label{eq6}
\end{eqnarray}

For square-wave stimulus, one should replace $\gamma $ by $4/T$ in
Eq.(\ref{eq6}).

Note that the HST occurs even for large additive noise ($\sigma
\gg F$). Indeed, we get from Eqs. (\ref{eq5}) and (\ref{eq6}),
using an asymptotic expression $\Phi \left( {a} \right) = 2a/\sqrt
{\pi}, \quad a \ll 1$, and taking into account that $V(F)$ is an
odd function:

\begin{eqnarray}
  &V = {\frac{{F}}{{\sigma} }}{\frac{{\pi \gamma} }{{2}}}A,
\nonumber \\
  &A = {\frac{{2\Delta} }{{\sqrt {\pi}  \left( {\Delta ^{2} - z_{c}^{2}}
\right)}}}\left( {\sqrt {\Delta + z_{c}}  + \sqrt {\Delta - z_{c}}
} \right).
\label{eq7}
\end{eqnarray}

Therefore, since $\sigma \sim  10^{ - n}$, the macroscopic
transport still exists.

Numerical integration of Eqs.(\ref{eq1}) was performed using an
Euler algorithm with the time step $\Delta t=0.01$. Values of $V$
were obtained by averaging over 100 runs of the model with random
initial conditions. The duration of one run was about $10^{7}$
except the cases when the value of $\varphi $ grew during that
time in such extent that the finite computer precision (20
significant digits) did not allow system to respond to an
ultrasmall signal $F = 10^{ - 11}$. In these cases the same
statistics was obtained using more runs with shorter duration. In
all simulations we take $z_{c}=0$ and $\sigma = 0$.

Fig.2 demonstrates the dependence of $V$ on noise correlation time
$\gamma $ for fixed noise amplitude $\Delta =10$ and various
control stimuli. The results for deterministic one are in
excellent agreement with theory. It is seen that HST disappears
for stimulus frequency greater than that determined by
(\ref{eq4}). The results for adiabatic stochastic stimulus
demonstrate a good agreement with second Eq.(\ref{eq6}) that is
shown in Fig.2 with dashed line. One can see also that the
transport occurs for high $\gamma $ values, though in a less
extent, therefore displaying a resonance on noise correlation
time. It is interesting to note that near the adiabatic boundary
(just left of the maximum) the deterministic stimulus induces HST
more effectively than the stochastic one.

Finally, Fig.3 shows the flux dependence on the amplitude of
additive noise $\sigma $. It is evident that the HST is robust to
the latter, since even for signal-to-nose ratio 0.01 the transport
remains macroscopic, decreasing as 1/$\sigma $, in accordance with
Eq.(\ref{eq7}).

To conclude, we note that in previous publications \cite{ours} we
reported a related phenomenon, the hypersensitivity to small
time-dependent signals in a double-well system that manifests
itself as large oscillations synchronized with signal. In the
present work we demonstrate that a system with periodic potential,
under effect of strong parametric, stochastic or deterministic,
stimulus exhibits hypersensitivity to an ultrasmall dc signal,
responding by macroscopic transport.

The work was supported by the RBRF Grant 99-02-17545, by the State
Program on Physics of Quantum and Wave Processes, Statistical
Physics Subprogram, Project VIII-3, and by the State Program on
Neutron Condensed State Research.

\bigskip
\newpage
Figure captions

\bigskip

Fig.1. The potential $U_{\pm}  \left( {\varphi}  \right) + F$ for
``two-state'' stimulus (Eq.(\ref{eq2}))

\bigskip

Fig.2. The transport velocity $V$ vs. correlation time $\gamma $
for 1) dichotomous stochastic stimulus (squares), 2) Gaussian one,
obtained by Ornstein-Uhlenbeck process simulation (triangles), and
3) square wave (circles). The fixed amplitude $\Delta =10$.

\bigskip

Fig.3. The transport velocity $V$ vs. intensity of additive white
noise at $F=10^{-11}$.
  The control noise is dichotomous with $\gamma =0.02$
and $\Delta =10$. The line is drawn in accordance with
Eqs.(\ref{eq5}) and (\ref{eq6}).

\begin{figure}
 \label{Fig1}
 \end{figure}
\begin{figure}
 \label{Fig2}
 \end{figure}
\begin{figure}
 \label{Fig3}
 \end{figure}

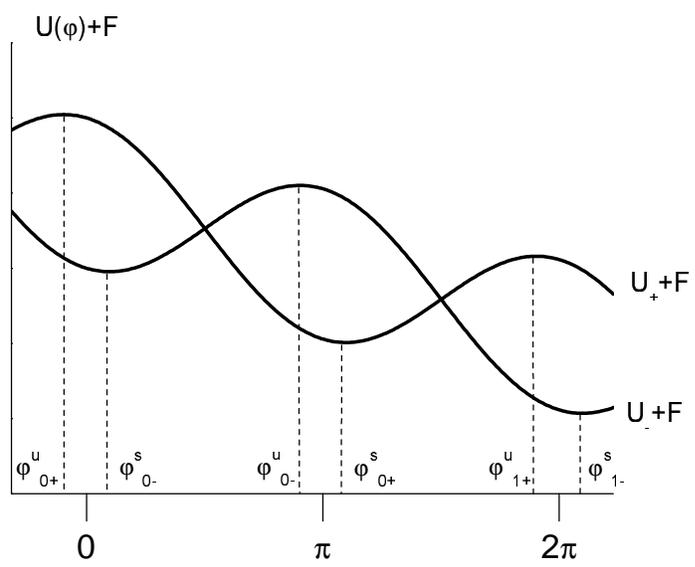
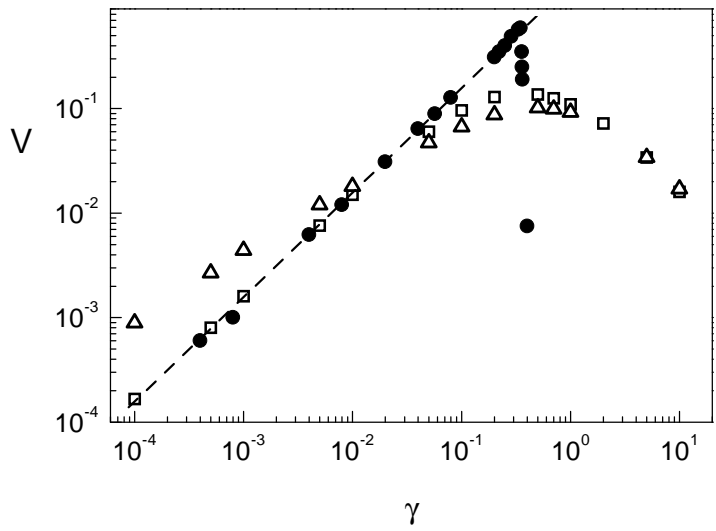
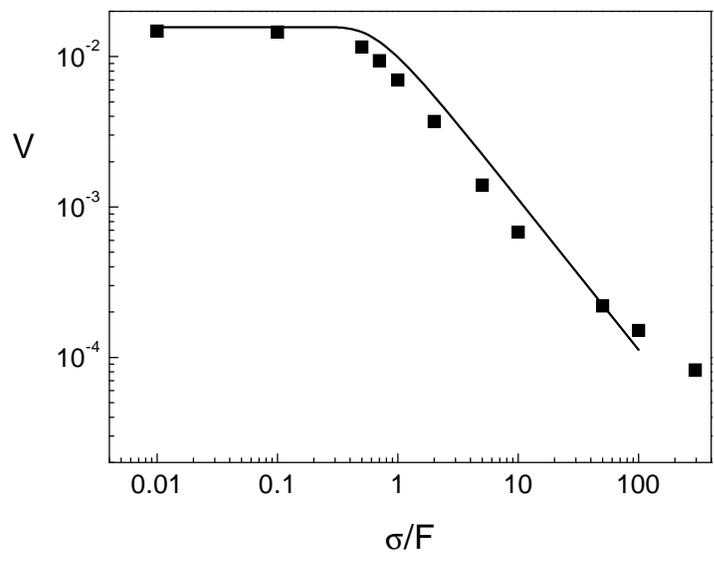
\end{document}